\documentclass[10pt,preprint]{aastex}
\newcommand{\kte}{kT$_{\rm e}$}
\newcommand{\ktbb}{kT$_{\rm BB}$}
\newcommand{\rbb}{R$_{\rm BB}$}
\newcommand{\ergs}{ergs s$^{-1}$}
\newcommand{\ergscm}{ergs s$^{-1}$ cm$^{-2}$}
\newcommand{\mdot}{$\dot{\rm M}$}

\newcommand{\ledd}{L$_{\rm Edd}$}
\newcommand{\comptt}{{\sc Comptt}}

\newcommand{\nh}{N$_{\rm H}$}

\slugcomment{Revised version, May 3rd, 2002}

\shorttitle{When an Atoll looks like a Z}
\shortauthors{Barret \& Olive.}
\def\fouru{4U $1705-44$}
\def\4U1705{4U $1705-44$}
\begin{document}
\title{A peculiar spectral state transition of 4U1705--44: \\ when an
Atoll looks like a Z}
\author{Didier Barret \& Jean-Fran\c{c}ois Olive} \affil{Centre d'Etude
Spatiale des Rayonnements, CNRS/UPS, \\ 9 Avenue du Colonel Roche, 31028 Toulouse
Cedex 04, France}
\email{Didier.Barret@cesr.fr}
%
%
\begin{abstract}
We report on a clear spectral state transition of the neutron star
  low-mass X-ray binary \fouru~observed by the Rossi X-ray Timing
  Explorer. In the X-ray color-color diagram (CCD), the source,
  classified as an Atoll, samples the upper parts of a Z, starting
  from and returning to the left bottom of the Z. We follow the path
of \fouru~on its color-color diagram, model its broad band
  X-ray/hard X-ray spectrum, and compute the Fourier power density
  spectrum of its X-ray variability.  The energy spectrum
 can be described as the sum of a dominating Comptonized component,
  a blackbody and a 6.4 keV iron line.  During most observations,
  \fouru~displays strong band limited noise with an integrated
  fractional RMS varying from $\sim 10$ to $\sim 20$\%. The spectral  transitions of \fouru~are shown to be primarily associated with changes in the temperature of the Comptonizing electrons.  During the soft to hard transition, the source luminosity decreased from $\sim 2.1$ to $\sim 0.7 \times 10^{37}$ \ergs, whereas the hard to soft transition took place at higher luminosities between $\sim 2.5$~and $\sim 3.1 \times 10^{37}$\ergs. In the hard state (top branch of the Z), the source evolves from left to right on the CCD, while its luminosity smoothly increases. Along this branch, the increase in the soft color is related to a smooth increase of the blackbody temperature, while the electron temperature remained remarkably constant ($\sim 13$ keV). 

Our observations can be interpreted in the framework of a model made of a truncated accretion disk of varying inner radius and an inner flow merging smoothly with the neutron star boundary layer. The spectral evolution could be driven by changes in the truncation radius of the disk; e.g. the soft to hard transition could be caused by the disk moving outwards. If this model is correct, then our data show that the disk truncation radius is not set by the instantaneous mass accretion rate, as derived from the source bolometric luminosity.

Comparing the power density spectra of \fouru~and Z sources when they occupy similar branches of the Z, we  show that the most significant difference is on the diagonal branch, on which the power density spectra of \fouru~remain similar to the ones measured on the top branch of the Z (hard state).
\end{abstract}


\section{Introduction}

\begin{figure}[t]
\begin{center}
\epsscale{1.0}
\plotone{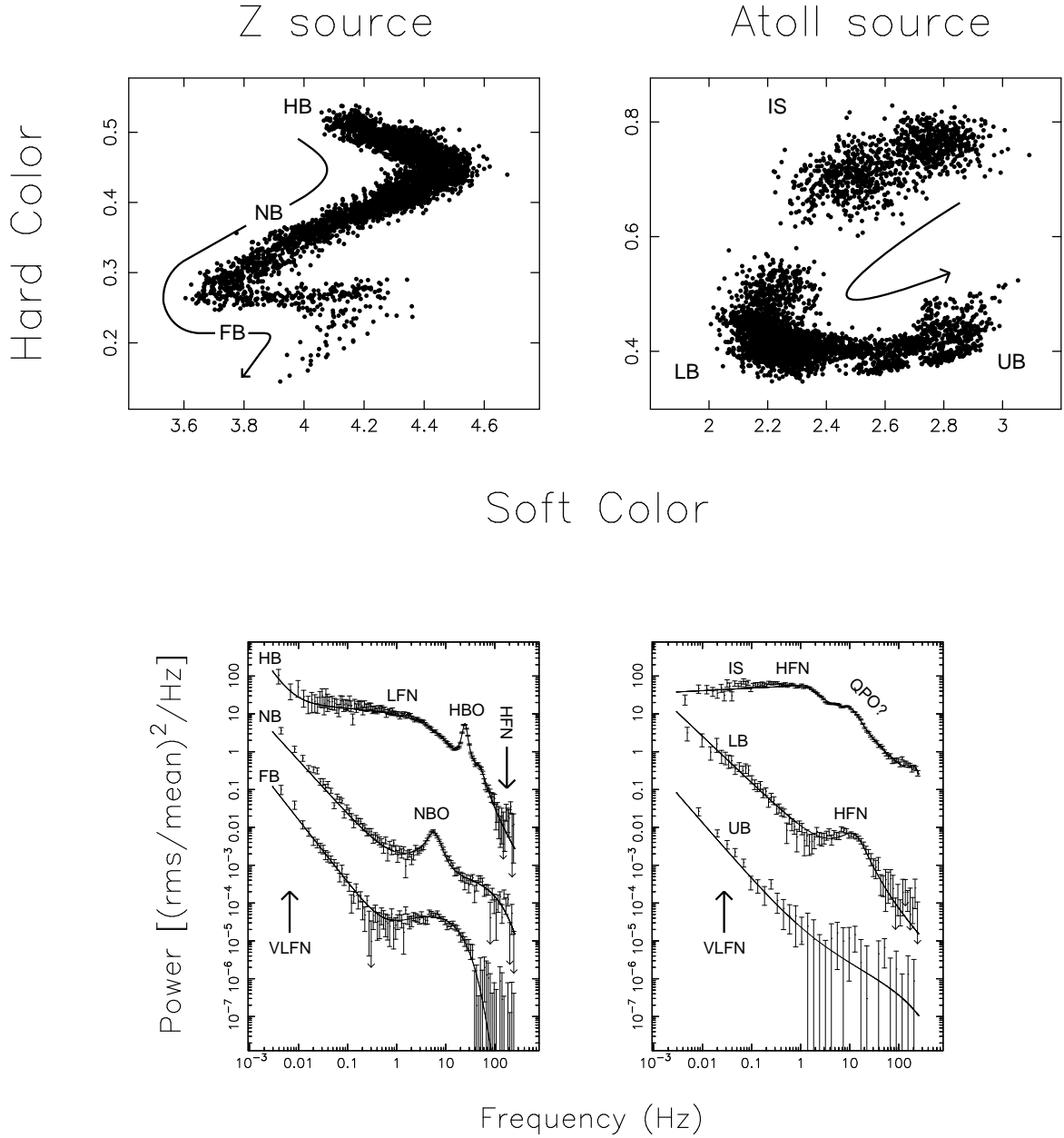}
\caption{X-ray color-color diagrams and Fourier power density 
  spectra of typical Z (left) and Atoll (right) sources taken from the
  review article of Wijnands (2001). For Atolls, IS means island
  state, LB lower banana, and UB upper banana. For Z, HB means
  horizontal branch, NB normal branch, FB flaring branch. The arrow is
  thought to give the sense of accretion rate (\mdot) variations along the diagram.
  \label{fig1_bo}}
\end{center}
\end{figure}

Low-Mass X-ray Binaries (LMXBs) with neutron star primary have been
divided in two classes; the so-called Atoll and Z sources (Hasinger \& van der Klis 1989).  This classification derives from the different
shape of their X-ray color-color diagrams (CCD).  In CCD, Z sources display a
Z-like track, whereas the Atolls display a more curved shaped track
(M{\' e}ndez 1999) (C-like) in which an ``island'' and a
horizontally elongated ``banana'' states can be identified (see Fig.
\ref{fig1_bo} taken from Wijnands 2001). Different positions on the
CCD are associated with different power density spectra of their X-ray variability for both
classes of sources (see also Fig. \ref{fig1_bo}). The position of the source on the CCD is thought
to be a good indicator of the mass accretion rate (\mdot), which in
the case of Atoll would increase from the island to the banana states
(see the arrows in Fig.  \ref{fig1_bo}).  Many properties of these
systems have been shown to correlate well with the source position in
the CCD (M{\'e}ndez 1999).  The differences between Atolls and Z were
thought to reflect differences in the mass accretion rate and neutron
star magnetic fields; where Z sources would have on average larger
magnetic fields and larger mass accretion rates in agreement with
their larger luminosities (Lamb et al. 1985; Hasinger \& van der Klis 1989).

However, recently, two groups have independently shown that, when
considering large amounts of data, sampling wider ranges of source
states, Atolls CCDs tend to resemble those of Z sources. This is
certainly the case for \fouru, 4U1608-522 and Aql X-1 (Gierli{\'n}ski
\& Done 2002a; Muno, Remillard \& Chakrabarti 2002). It has thus been
argued that the previous distinction between Z and Atolls could be an
artifact of incomplete sampling of the source intensity states, further
leading to the conclusion that Z and Atolls are actually responding
similarly to a varying mass accretion rate.

In this paper, we report on a clear spectral state transition of
\fouru~observed during consecutive RXTE observations.  During these
observations, in its CCD the source sampled the top and diagonal branches of the
Z reported by Gierli{\'n}ski \& Done (2002a) and Muno et al. (2002).  We extend previous work on the source by
following its path on the CCD, modeling its broad band energy spectrum
and computing the Fourier power density spectrum of its variability
along the whole transition.  This enables us to determine which
spectral parameters drive the changes in X-ray colors, to get some insights on the
accretion geometry changes associated with the spectral transitions, and finally compare the
general properties of \fouru~with those of Z sources when they
occupy similar branches of the Z on their CCD. 
We present the RXTE observations and the results of our spectral and
timing analysis of the persistent emission of \fouru~in section
\ref{observations}, and discuss our results in section
\ref{discussion}.

\section{Observations and results}
\label{observations}

\begin{figure}[t]
\begin{center}
\epsscale{1.0}
\plotone{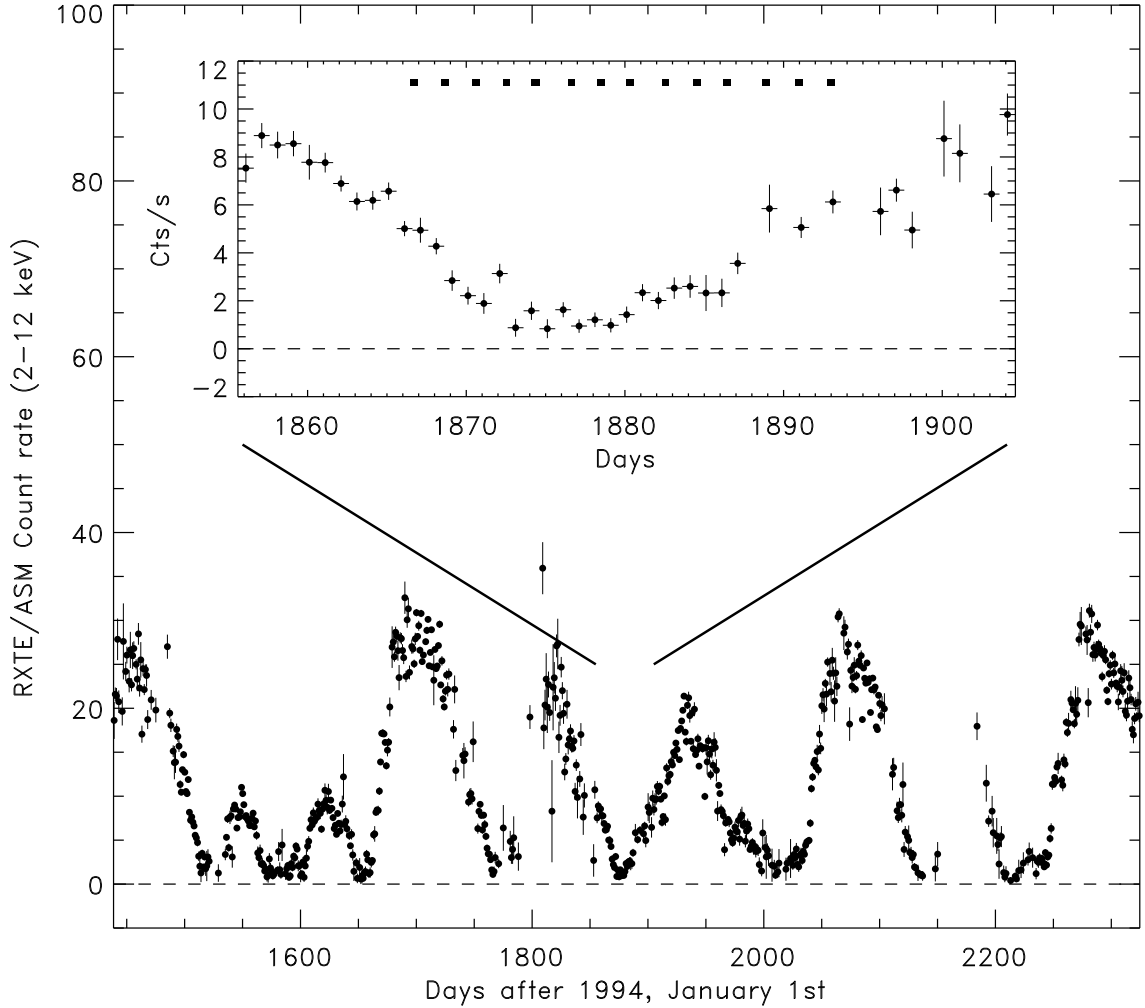}
\caption{Long term RXTE/ASM 2-12 keV light curves. For information about
the ASM, see Levine et al. (1996). The data were retrieved from the
HEASARC public database. The upper panel represents a zoom around the time
period during which the RXTE pointed observations were performed. The time
of the pointed observations are shown at the top of the plot with square
dots. The decay and the rise of the event were perfectly sampled by the
pointed observations. The source intensity ranges from $\sim 15$ mCrab up
to $\sim 100$ mCrab during our observations. Overall the source intensity
can reach $\sim 500$ mCrab in X-rays. \label{fig2_bo}}
\end{center}
\end{figure}

\fouru~is a classical LMXB (Liu, van Paradijs \& van den Heuvel 2001 and reference therein) characterized by large intensity variations, which make it an ideal
tool to study spectral state transitions in those systems.  The large
variability of the source is best illustrated by the long term RXTE
All Sky Monitor (ASM) light curve shown in Figure \ref{fig2_bo}.

Over the last few years, \fouru~has been observed many times by RXTE
(for a description of RXTE, see Bradt, Rothschild \& Swank 1993). One particular set
of observations (proposal number 40051 led by Dr.  Shuang Nan Zhang)
deserves some attention as it started while the source intensity was
slowly decaying (in February-March 1999) and then rising again to its
original level (see upper panel of Figure \ref{fig2_bo}).  A total of
14 snapshots were performed during this episode (see square dots in
Fig.  \ref{fig2_bo}). The RXTE observation log is given in Table
\ref{table1}. As can be seen, the observations span $\sim 30$ days,
with about 2 days between each.

\subsection{PCA and HEXTE light curves}

\begin{figure}[!t]
\begin{center}
\epsscale{1.0}
\plotone{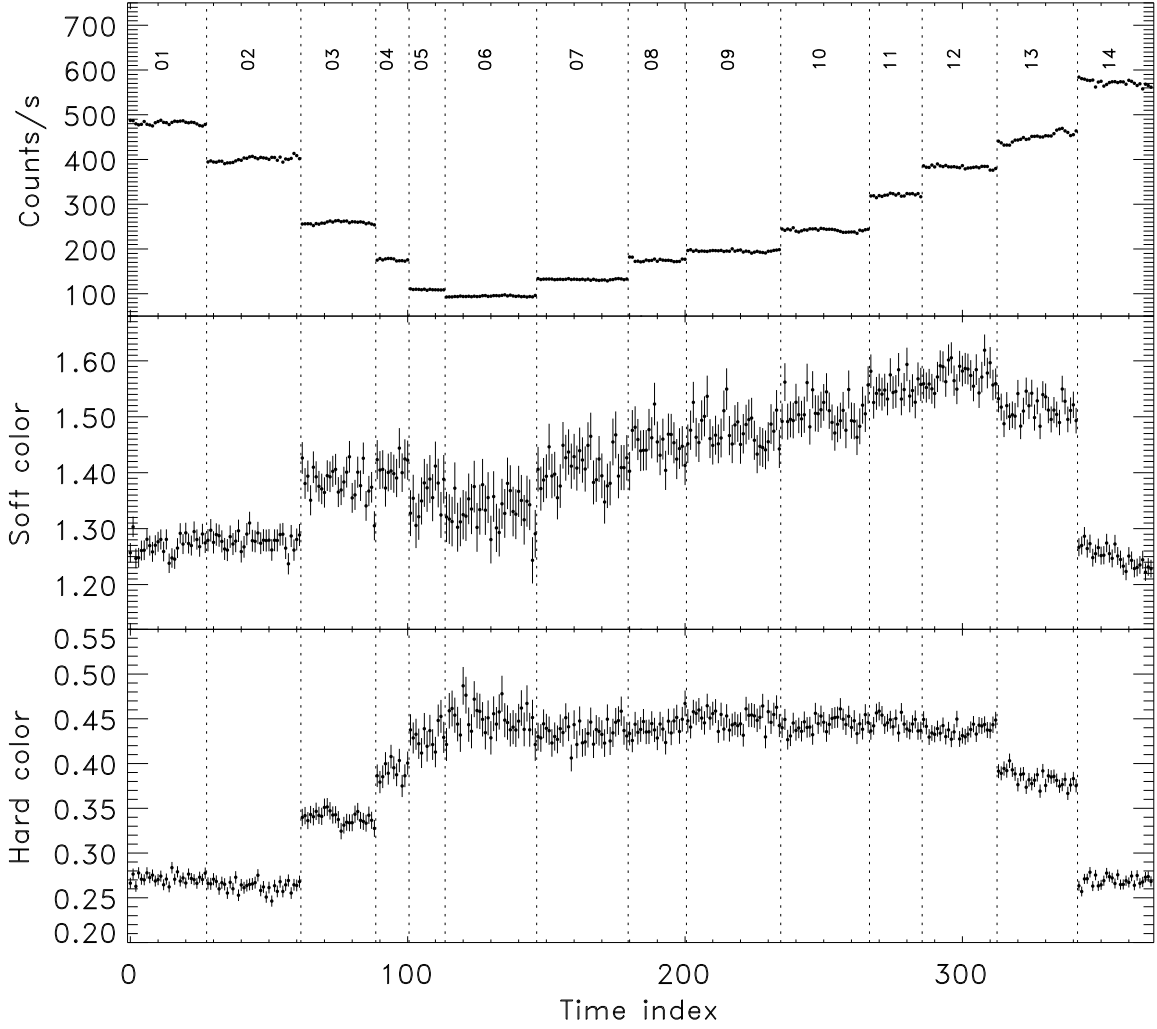}
\caption{The PCA background subtracted light curve of
  the \fouru~in the 3-16 keV band (top panel). The data are plotted
  versus a time index to avoid the gaps in time between observations.
  Each bin is 128 second long. Soft color (HR1, center panel) and hard
  color (HR2, bottom panel) as a fonction of time.
  \label{fig3_bo}}
\end{center}
\end{figure}

\begin{figure}[!t]
\begin{center}
\epsscale{1.0}
\plotone{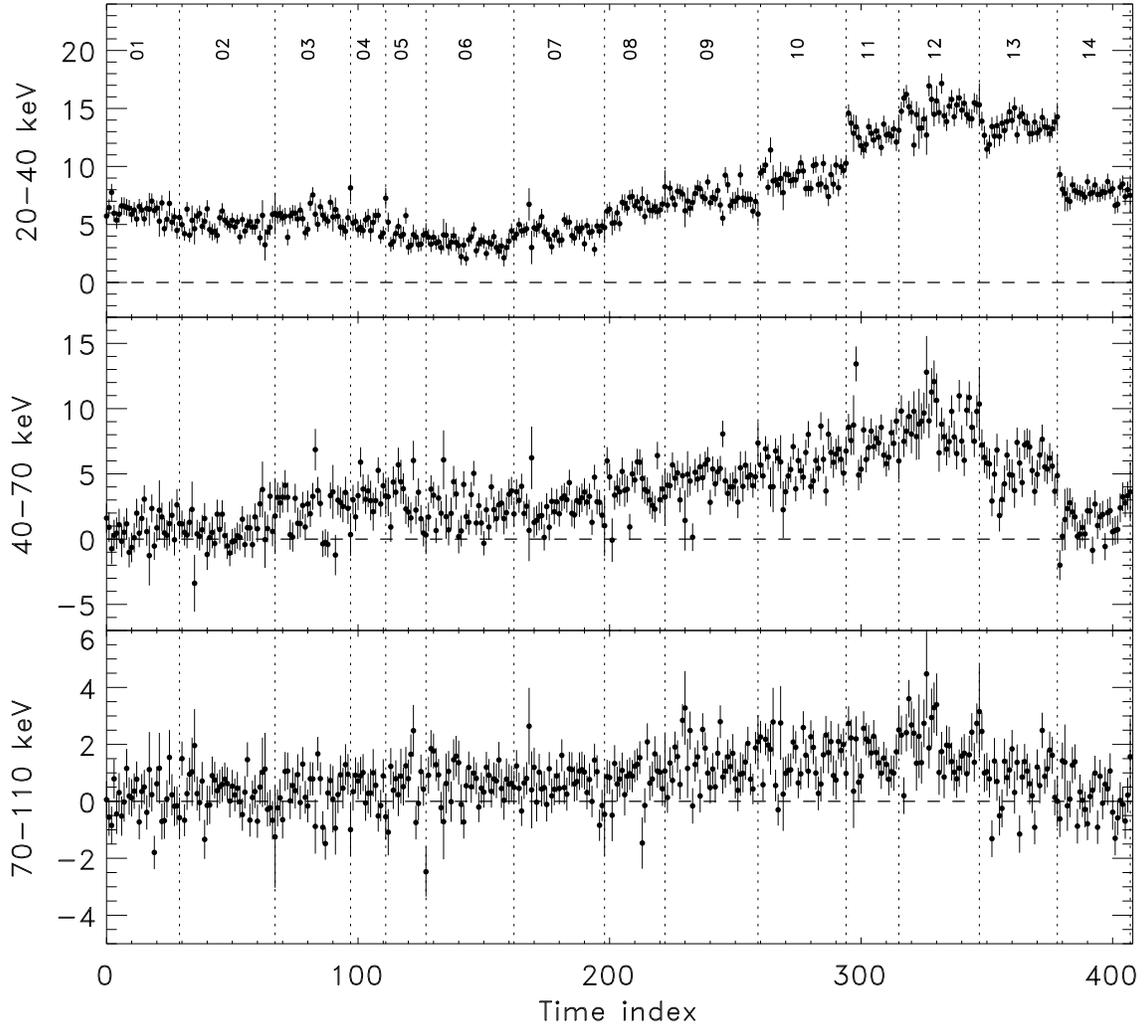}
\caption{HEXTE-0 background subtracted light curve of
  \fouru~as a function of time index in three consecutive energy
  bands. The bin time is 128 seconds. \fouru~is detected above the 3$\sigma$ level between 70 and $\sim 110$ keV in the observations 3
to 13.}
\label{fig4_bo}
\end{center}
\end{figure}

We analyzed the RXTE PCA and HEXTE data using FTOOLS 5.1 and following
standard recipes, using a customized version of REX 0.3.  PCA light
curves were extracted from PCA units 0, 1 and 2, in four consecutive
energy bands: 2.9-4.3 keV, 4.3-6.5 keV, 6.5-10.1 keV, and 10.1-16.2
keV (REX channels 8-11, 12-17, 18-27, 28-44). We filtered the data
using standard criteria: Earth elevation angle greater than 10
degrees, pointing offset less than 0.02 degrees, time since the peak
of the last SAA passage greater than 30 minutes, electron
contamination less than 0.1.  The background of the PCA has been
estimated using {\it pcabackest} version 2.1e.  The PCA light curves
from the 14 pointings are shown in the top panel of Figure
\ref{fig3_bo} (hereafter called observations 1 to 14).  Two bursts
detected during observations 4 and 11 were filtered out.  We show the
HEXTE-0 light curves in Figure \ref{fig4_bo} in three consecutive
energy bands: 20-40 keV, 40-70 keV, 70-110 keV. The
source is detected above the 3$\sigma$ level between 70 and $\sim 110$ keV in the observations 3
to 13.

\subsection{Color-color diagram}

We followed the spectral evolution of the source during the
observations by computing hardness ratios: the soft color (HR1) is
defined as the ratio between the 4.3-6.5 keV counts and 2.9-4.3 keV
counts, and the hard color (HR2) as the ratio between the 10.1-16.2
keV counts and 6.5-10.1 keV counts.  The time evolution of the soft
and hard colors is displayed in the lower panels of Figure
\ref{fig3_bo}.  The most spectacular spectral transitions occurred
between observations 2 and 4 (in 4 days), and in a reverse way between
observations 12 and 14 (again in 4 days).  It is remarkable that after the first
transition, a slight decrease of HR1 can be observed simultaneously
with an increase of HR2 which reached a maximum in
observation 6.  Then HR2 remained remarkably constant whereas HR1 kept
increasing very progressively (observations 7 to 12, 12 days).  It is worth
noting that while HR2 remained constant, the hard X-ray flux observed
by HEXTE continued to increase (see Fig. \ref{fig4_bo}).  This indicates that the spectral
shape between 6.5-16.0 keV remained constant, only the hard X-ray flux
increased.  In observation 14, \fouru~had returned to a spectral state
very similar to the one of observations 1 and 2. Taking the mean colors during each observation, the path followed by the source on its CCD during the transition is shown on Fig. \ref{fig5_bo}. 

To complete the lower part of Z, we reanalyzed some archival data from proposal number P20074 (led by Dr.  Phil Kaaret), in which the source was observed at much higher
intensities (June-September 1997).  For that purpose, we computed the
hardness ratio in similar energy bands, and correcting for the shift
in gain of the PCA between the two observations (both are epoch 3
data). The X-ray colors are very sensitive to the exact values of
  the channel energy boundaries over which they are computed. For the two observations, we computed the corresponding PCA energy responses to get the channel energy boundaries. We fitted with a polynomial function the count spectrum (normalized in counts/s/keV) of observation P20074. We then computed the expected number of counts by
  numerically integrating the fitted polynomial function over the 
  energy boundaries of the observation P40051, which is the reference point for the X-ray colors. This is the exact treatment
  which naturally accounts for the shape of the source spectrum and
  gain shifts. Gain shift induced color variations within the data of
  observation P40051 (spanning one month) are less than 1\% and are therefore negligible. Residual variations due to the approximations used in the PCA response generation as reported in (Gierli{\'n}ski \& Done 2002a) are also negligible.
The CCD shown in Fig. \ref{fig6_bo} combines the two data sets and
reveals the Z shape reported previously by Gierli{\'n}ski \& Done
(2002a) and Muno et al. (2002). The data
taken from observation P20074 sampled the bottom branch of the Z, during which
the averaged unabsorbed 0.1--200 keV flux was $1.3\times
10^{-8}$\ergscm, corresponding to a bolometric source luminosity of
$\sim 8.8\times 10^{37}$\ergs~at the distance of 7.4 kpc (Haberl \&
Titarchuk 1995). This is just about a factor of 10 larger than the minimum luminosity of the observations of proposal P40051. This means that \fouru~samples the Z shape while its luminosity changes by only a factor of 10. Hence one might be able to observe a Z, even from those sources characterized by moderate intensity variations and not only from those sources exhibiting very large (factor 100--1000) intensity variations (Gierli{\'n}ski \& Done 2002a; Muno et al. 2002). 

\subsection{Spectral analysis}
\label{spectral_analysis}
We extracted PCA (units 0,
1, 2, top layer only) and HEXTE spectra for each observation.  For the
PCA response matrix generation we used {\it pcarsp.v7.11}.  To
account for uncertainties in the PCA response (see for example Tomsick, Corbel \& Kaaret 2001), a systematic error of 0.5\% has been added to the PCA
count spectra corresponding to the four brightest observations. The
PCA spectra were fitted between 3.0 and 25.0 keV whereas for HEXTE,
the data were fitted up to $\sim 30$ keV for the soft spectra and up
to $\sim 100$ keV for the hard spectra. For all observations, the PCA and HEXTE (0 and 1) spectra were fitted jointly, leaving the relative normalization between spectra as free
parameters of the fit. 

With the spectral analysis of the persistent emission, we wish to determine which parameters drive the changes observed in the CCD. For this purpose, we tried to fit the whole data set with several physical models, while trying to keep the number of free parameters as low as possible. The most simple model which provides a good fit to the data is made of the sum of a thermal comptonized component modeled in XSPEC by \comptt~(Titarchuk 1994) (a spherical geometry was assumed but a disk geometry works equally well), a soft component which we modeled by a single temperature blackbody and an iron line whose energy was fixed at 6.4 keV. Similar models have been shown to apply to several Atoll sources in similar states (e.g. Barret et al. 2000). Leaving the line
energy, as a free parameter yields an average value very close to and
consistent with 6.4 keV.  The line width ($\sigma_{\rm 6.4 keV}$) was
found to be consistent with being constant and narrow, with a mean
value of $\sim 0.7$ keV. Such a line could be produced through the irradiation of the accretion disk by the central X-ray source (e.g. Barret et al. 2000).

The column density (\nh) towards the source is not accurately known
and cannot be well constrained by the PCA whose energy threshold for
the fit is $\sim 3$ keV (see e.g. Langmeier et al.  1987).  It was
initially left free in order to determine its average value.  No
significant \nh~variations were found within the observation data set.
Therefore, its value was frozen at the average ($2.4\times 10^{22}$
cm$^{-2}$) measured by the PCA. This value is consistent with the one reported by Vrtilek et al. (1991). Fixing the absorption makes the fitting procedure more stable by avoiding the degeneracy between the absorption and blackbody parameters.

Similarly, during the first run we found that the seed photon
temperature derived from \comptt~was systematically lower than 1.0
keV, and also poorly constrained.  In the fits, its value was frozen
at the mean observed during the first run (i.e. 0.4 keV).  Finally,
\fouru~is relatively faint for HEXTE, which means that for the lowest
statistics spectra, the electron temperature and the optical depth ($\tau$) of
the \comptt~model cannot be constrained simultaneously. This is the
case for observations 4 to 8. We checked, by computing hardness ratios
between (40-70 keV to 20-40 keV) in HEXTE, that the source did not
exhibit any significant spectral changes in the HEXTE energy range
between observation 4 to 10 (note that the constancy of the hard color as shown on Fig. \ref{fig3_bo} indicates that the spectral shape between 6 and 16 keV is constant as well). We set $\tau$ to 5.5 for observations 4 to 8, as this is the mean value derived from observations 9 and 10, during which it was the best constrained. Leaving $\tau$ as a free parameter yields consistent results within
error bars. The results of the spectral fitting are listed in Table
\ref{table2}.  Four representative spectra are shown in Figure
\ref{fig7_bo}. The way the spectral parameters and luminosities evolve
during the observations is shown in Fig \ref{fig8_bo} and Fig.
\ref{fig9_bo}.

Finally, beside the above model, is worth noting that for observations 1, 2 and 14 (high luminosities), another simple physical model fit the data as well. In this model, the seed photon temperature is left free and the soft component is fitted by a multicolor disk blackbody. The parameters of the comptonized component (electron temperature and optical depth) do not change significantly with respect to the model described above. In all three observations, the seed photon temperature derived is $\sim 1.1$ keV, the inner disk temperature is $\sim 0.6-0.7$ keV, and the projected inner disk radius ($R_{\rm in} \sqrt{\cos\theta}$) is $\sim 20\pm5$ km (at the distance of \fouru). Such a model has also been shown to apply to Atoll sources when observed in similar spectral states (e.g. 4U1608-52, Gierli{\'n}ski \& Done 2002b).

\subsection{Power Density Spectra}

Power density spectra were computed using E\_16us\_64M\_0\_1s data in
the 0.015-2048 Hz range between 3 and 30 keV, in the exact same time
intervals as the energy spectra. The spectra of observations 7, 8, and
9, 10 are very similar and were grouped together. The Poisson counting noise level
estimated between 1300 and 2048 Hz was subtracted. The 12 power
density spectra so computed together with the integrated RMS amplitude
are shown in Figure \ref{fig10_bo}.

Without entering into the details of the shape of the power density
spectra, there are several noticeable features worth mentioning. All
power density spectra are characterized by the presence of a strong
band-limited noise component. This noise is made of up to 3 broad
features whose characteristic frequencies evolve during the
observations. Noise is detected up to $\ge 500$ Hz between
observations 7 to 13, during which the total integrated RMS amplitude
is the largest (15--18\%). This high frequency component has been established in several Atoll sources, and proposed as a possible way to distinguish between black hole candidates and neutron star systems (Sunyaev \& Revnivtsev 2000; see however Belloni, Psaltis \& van der Klis 2002). Between observations 1 and 2, whereas the
energy spectrum remained unchanged despite a decrease of $\sim 15$\%
in count rate, the power density spectra of observation 2 reveals the
presence of an additional low-frequency component below 1 Hz.  This
shows that, there is not a one-to-one relationship between the energy
spectrum and the variability of the source, and there are processes
contributing to or modifying the power density spectra without
affecting the shape of the energy spectrum. The power density spectra
of observations 1 and 14 are very similar in shape but differ in RMS
values, due to the presence of a $\sim 750$ Hz Quasi-Periodic
Oscillation (QPO) in the latter observation (it is barely visible in
Fig.  \ref{fig10_bo}). A detailed modeling of the power density
spectra with multi-Lorentzians (e.g. as in van Straaten et al.  2000;
Belloni et al. 2002), and the evolution of the characteristic
parameters of these Lorentzians as a function of colors, spectral
parameters and fluxes will be reported elsewhere.

\section{Discussion}
\label{discussion}
We carried out a spectral and timing study of \fouru~during a well sampled spectral state transition. We discuss below the results so obtained, emphasizing the spectral parameters driving the changes in
X-ray colors, the path followed by the source on its CCD during the transition, and the implications of our observations in the framework of an accretion model proposed for Atoll sources. Then, we briefly comment on the remaining differences between Z and Atolls.
\subsection{Spectral transitions in \fouru}

The energy spectrum of \fouru~is accurately fitted by the sum of a Comptonized component, a blackbody, and a 6.4 keV iron line. Within this model, the two spectral parameters which vary the most during the
observations are the blackbody temperature (\ktbb) and the electron temperature of the Comptonizing cloud (\kte). The spectral transitions soft to hard and later hard to soft (associated in the CCD with a decrease/increase of the hard color) are clearly related to an increase and then a decrease of \kte. The good correlation between the hard color and \kte~is shown in Fig. \ref{fig11_bo}. In the hard state (at constant hard color), changes in the soft color are driven by changes in \ktbb~(see Fig. \ref{fig11_bo}).

During our observations, the bolometric source luminosity computed by
extrapolating the fitted model from 0.1 to 200 keV ranged from $6.9
\times 10^{36}$ to $3.1 \times 10^{37}$ \ergs~(see Fig \ref{fig9_bo}).
The Comptonized component is always carrying most of the luminosity
(up to $\sim 95$\%). The fraction of luminosity in the blackbody is
the largest in observations 1, 2, and 14, when it reaches $\sim
20$\%. We showed that a significant fraction of the X-ray luminosity
can be carried out (up to 30\%) in hard X-rays (20-200 keV, Fig.
\ref{fig9_bo}). This is typical of Atoll sources in the low luminosity regime (e.g. Barret et al. 2000). The X-ray count rate is a good measure of the bolometric source
luminosity for soft spectra but can underestimate the luminosity of
the source by up to $\sim 30\%$ for hard spectra (see the
parameter $\eta$ in Table \ref{table2}).

\begin{figure}[!t]
\begin{center}
\epsscale{1.0}
\plotone{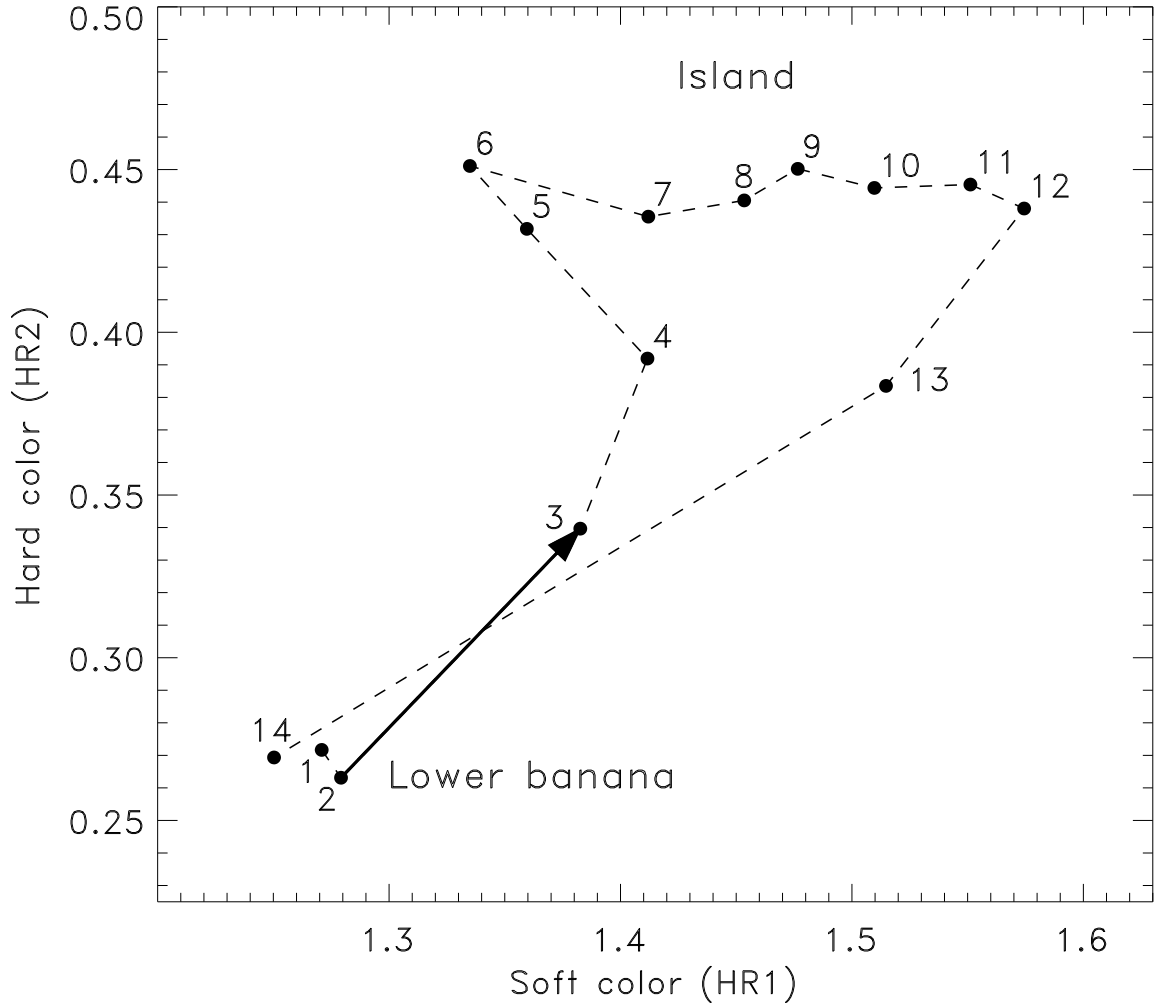}
\caption{The path followed by \fouru~on the color-color  diagram
  during the spectral transition reported here (see also Fig.
  \ref{fig6_bo}). Note that the source moved from the left to the
  right in the island state.  \label{fig5_bo}}
\end{center}
\end{figure}

\begin{figure}[t]
\begin{center}
\epsscale{1.0}
\plotone{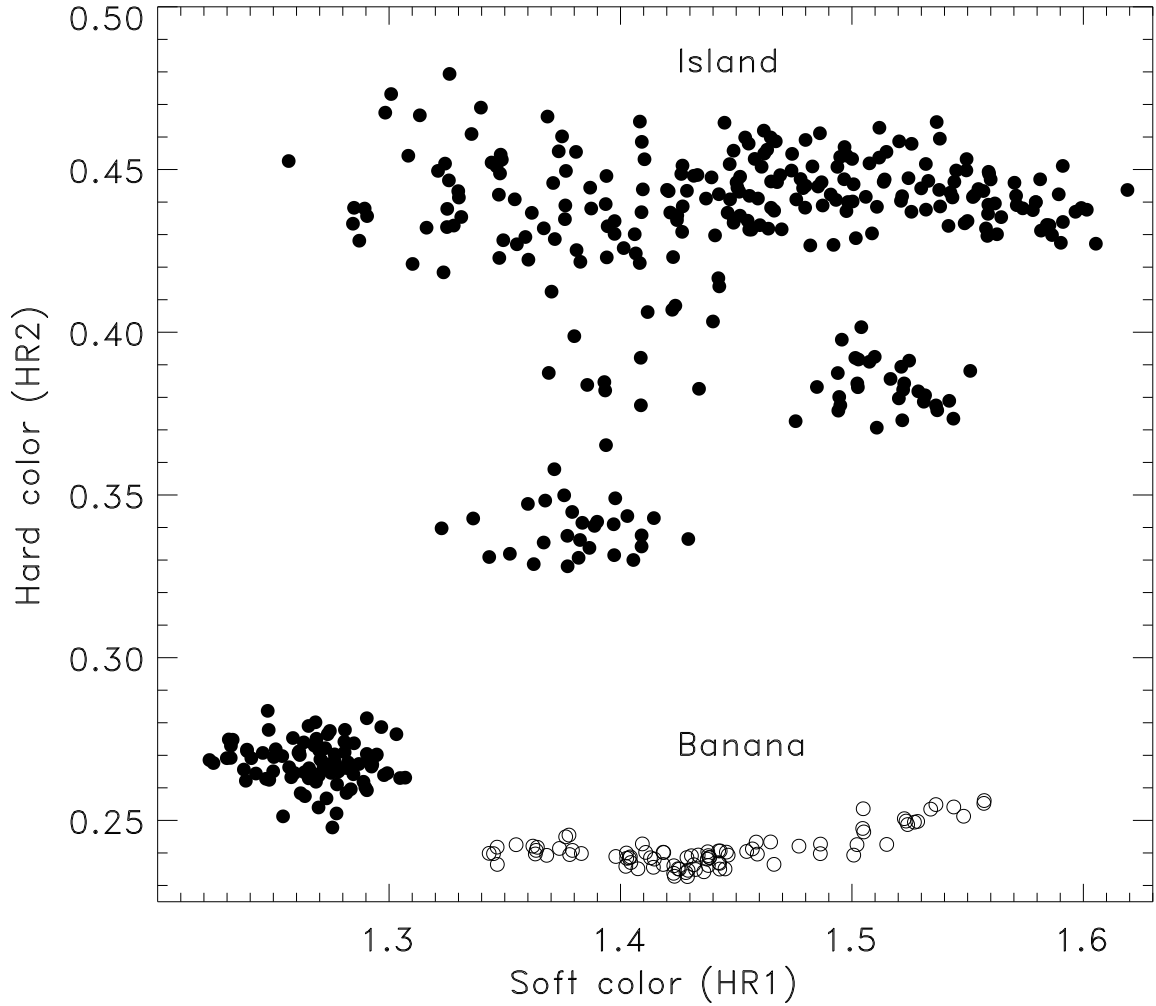}
\caption{X-ray color-color diagram of \fouru~combining data from the
  present observations (filled circles) and those from observation
  P20074 (empty circles). Our observations sample the island and lower
  banana states of \fouru~as well as the transition between the two
  states. The Z shape previously reported from the source is clearly
  visible. \label{fig6_bo}}
\end{center}
\end{figure}

\begin{figure*}[ht]

\begin{center}
\epsscale{1.0}
\plotone{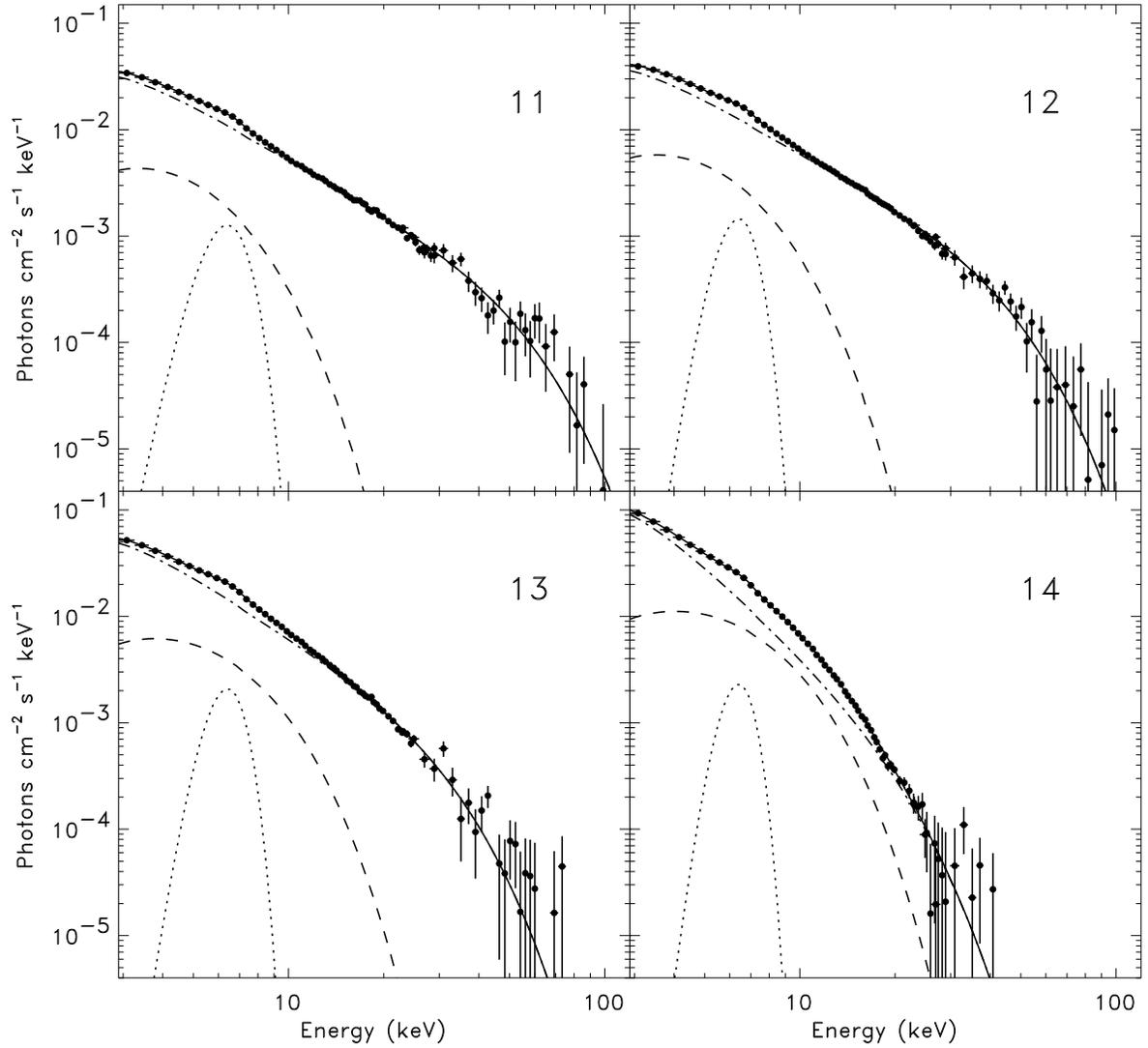}
\caption{Four unfolded spectra of \fouru~corresponding to observations 11
  to 14 (transition hard to soft). The dot line is the gaussian 6.4
  keV line, the dashed line is the blackbody component, and the
  dash-dot line is the Comptonized component.  Both the PCA and
  HEXTE-0 spectra are shown. The relative normalization between the
  PCA and HEXTE-0 spectra has been corrected for. The progressive
  softening of the spectrum is obvious (decrease in the electron
  temperature and increase in the blackbody temperature).
  \label{fig7_bo}}

\end{center}
\end{figure*}

\begin{figure}[t]

\begin{center}
\epsscale{1.0}
\plotone{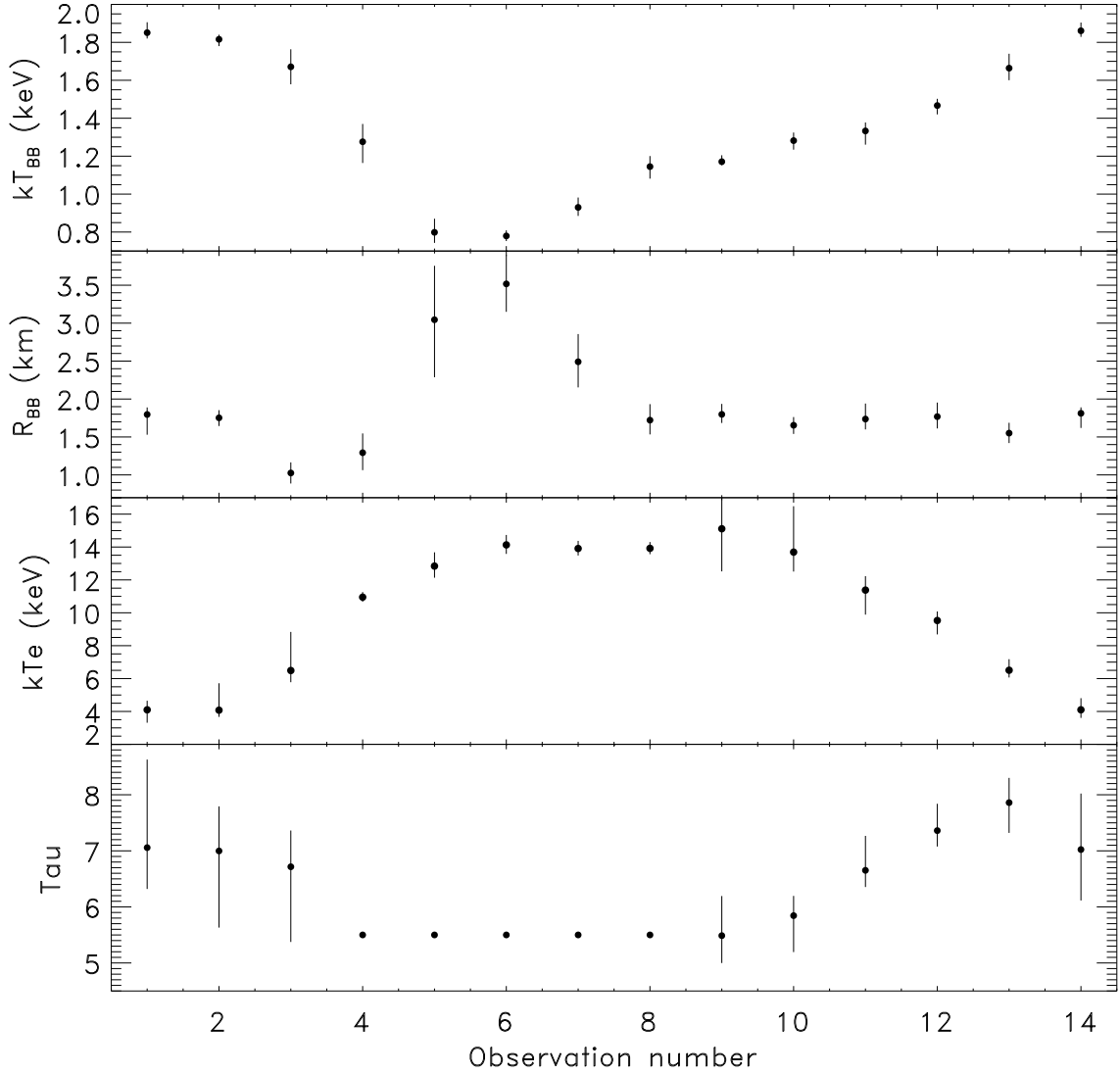}
\caption{Spectral parameter evolution of \fouru~along the observations.
  The X axis represents the observation number. The parameters are
  listed in Table \ref{table2}.  For observations 4 to 8, the optical
  depth has been frozen to 5.5; the mean value observed in
  observations 9 and 10.  This allows a better determination of
  \kte~in the \comptt~model.}\label{fig8_bo}

\end{center}
\end{figure}

\begin{figure}[t]
\begin{center}
\epsscale{1.0}
\plotone{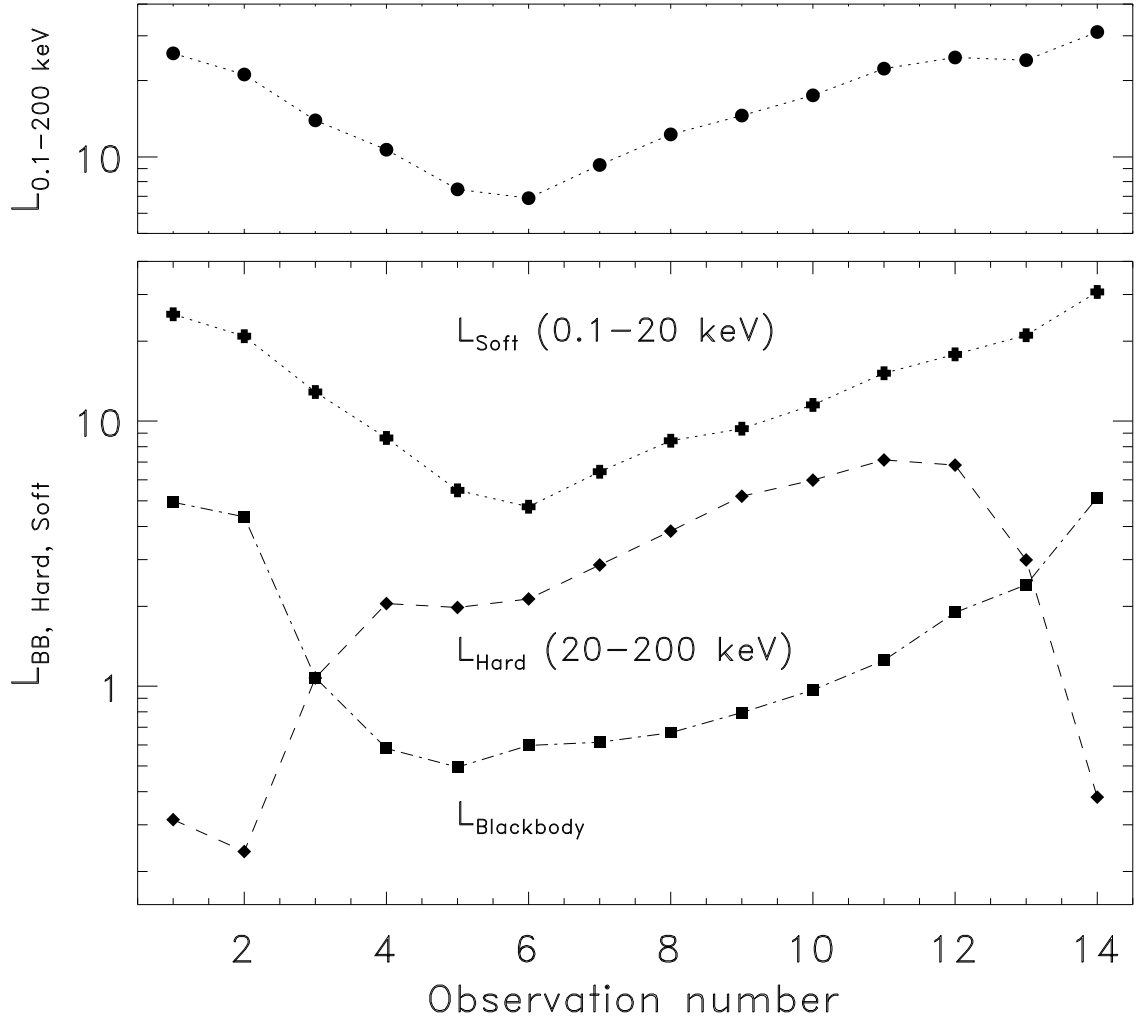}
\caption{The 0.1-200 keV total luminosity in units of $10^{36}$
  \ergs~ of \fouru~along the observations (top panel). Note the slight
  decrease of the source luminosity at observation 13. From bottom to
  top, the lower panel shows the blackbody (L$_{\rm Blackbody}$,
  filled squares), the hard (L$_{\rm Hard}$, 20-200 keV, filled
  diamonds) and the soft (L$_{\rm Soft}$, 0.1-20 keV, filled crosses)
  luminosities.  The spectral transitions can be visualized as a
  strong decrease/increase of the blackbody flux together with an
  increase/decrease of the hard X-ray flux.}
\label{fig9_bo}
\end{center}
\end{figure}

The present observations demonstrate that in the hard state, the source evolves from left to right in the CCD while the mass accretion rate (inferred from the bolometric source luminosity) increases (Fig \ref{fig5_bo}). This has already been shown by Gierli{\'n}ski \& Done (2002a) and is different from what has been assumed in the past (see Fig \ref{fig1_bo}). When moving from and to the bottom of the Z, the source followed different paths, as shown in  Fig. \ref{fig5_bo}. Between observations 1 and 6 during which the soft to hard spectral transition occurred, the bolometric luminosity decreased by about a factor of $\sim 3$; from  $\sim 2.1$ to $\sim 0.7 \times 10^{37}$ \ergs. On the other hand, during the opposite transition (hard to soft) between observations 12 and 14, the source luminosity was higher and increased only by a factor of $\sim 1.2$ (from $\sim 2.5$~to $\sim 3.1 \times 10^{37}$\ergs). In addition, during observation 13, the bolometric luminosity of \fouru~stopped increasing (it even decreases slightly, Fig. \ref{fig9_bo}). This observation is characterized by a strong decrease of the hard X-ray flux, whereas the soft X-ray flux continues to increase smoothly. This effect has been observed in other sources (Aql X-1, 4U1608-52) and led to the interesting speculation that it could be associated with the formation of a jet (Gierli{\'n}ski \& Done 2002a).  The source luminosity would stop increasing because some energy would be dissipated as kinetic energy in the jet instead of being radiated. So far, \fouru~has not yet been detected at radio wavelengths (Fender \& Hendry 2000), but radio observations during these episodes of variability would be worthwhile.

The X-ray spectra of Atoll sources are generally explained in the framework a model made of a truncated accretion disk with a hot inner flow which merges smoothly with the boundary layer (Barret et al. 2000; Barret 2001; see also Done 2002). Conduction of heat between the hot inner flow and a cold standard disk leading to the evaporation of the disk has been proposed as a mechanism for the transition (R{\' o}{\. z}a{\'  n}ska \& Czerny 2000). The truncation radius of the disk is the critical parameter of this model. This is because the disk represents a powerful source of cooling (in addition to the neutron star) for the comptonization; the closer the disk gets to the neutron star, the more effective it cools the inner flow (e.g. Done 2002). The spectral evolution of \fouru~presented here can be interpreted in the framework of this model, involving a varying disk truncation radius. 

At the beginning of our observations (1 and 2), the truncation radius is small; the disk is close to the neutron star surface. Its temperature is high enough to be detectable in the PCA band pass. At the same time, the inner flow is cool and the relatively large optical thickness of the inner flow prevents from seeing the neutron star surface. The soft component is therefore likely to arise from the disk. This would favor the spectral model in which the soft component is fitted with a multi-color disk blackbody (see section \ref{spectral_analysis}). 

Later on (from observation 3 to 6), the disk recedes (the truncation radius increases), and simultaneously the inner flow heats up while its optical depth decreases (the hard color increases). The soft to hard spectral transition might occur simultaneously with the disappearance of the disk emission and the appearance of the neutron star surface emission, resulting in the complex behaviour of the soft color, which first increased and then decreased. Unfortunately, the sensitivity and spectral resolution of the PCA do not allow to separate the disk and the neutron star surface emissions, and determine which one is present and dominates over the other.

From observations 6 to 12, the disk is still truncated at large radii and the soft emission remains dominated by the neutron star surface, and the corresponding temperature of the blackbody increases. This translates to an increase of the soft color, while the hard color remains constant. Again, the data does not allow to rule out a disk origin for the soft component, but the small radius measured for the blackbody component supports the idea that it comes from the neutron star. 

Later (between observations 12 and 13), the disk may get sufficiently close to the neutron star that its emission reappears in  the PCA band pass. Then, the disk contributes to the cooling of the inner flow, which becomes cooler and thicker. The soft and hard color decreases; this is the transition between the hard state to the soft state. In observation 14, the system is back to a state similar to observations 1 and 2. It has been shown that at even higher accretion rates (not explored with the present data), the temperature of the disk further increases, which leads to an increase of the soft color and thus explains the elongated shape of the soft state seen in the CCD (Gierli{\'n}ski  \& Done 2002b).  

Now the reasonable question to ask is which parameter sets the disk truncation radius? The data presented here demonstrate clearly that it cannot be the instantaneous mass accretion rate inferred from the bolometric source luminosity. This is best illustrated by comparing observations 3 and 9 (also 2 and 11) which have very similar inferred accretion rates and very different spectral shapes. One possibility could be that the truncation radius depends at some level on the average accretion rate over timescales of days, similarly to what has been proposed by van der Klis (2001) to explain the "parallel tracks" phenomenon in LMXBs. A detailed modeling of the Fourier power density spectra is underway to determine how the timing properties of \fouru~and their relation to its spectral evolution can better constrain this hypothesis.

\subsection{Atolls versus Z}
CCDs have been extensively used not only because they are sensitive to spectral variations, but also because different power density spectra are associated with different positions in the CCD. The similarities in the CCD of \fouru~and Z sources make relevant the comparison of the power density spectra of \fouru~and Z sources.

\begin{figure*}[tb]

\begin{center}
\epsscale{1.0}
\plotone{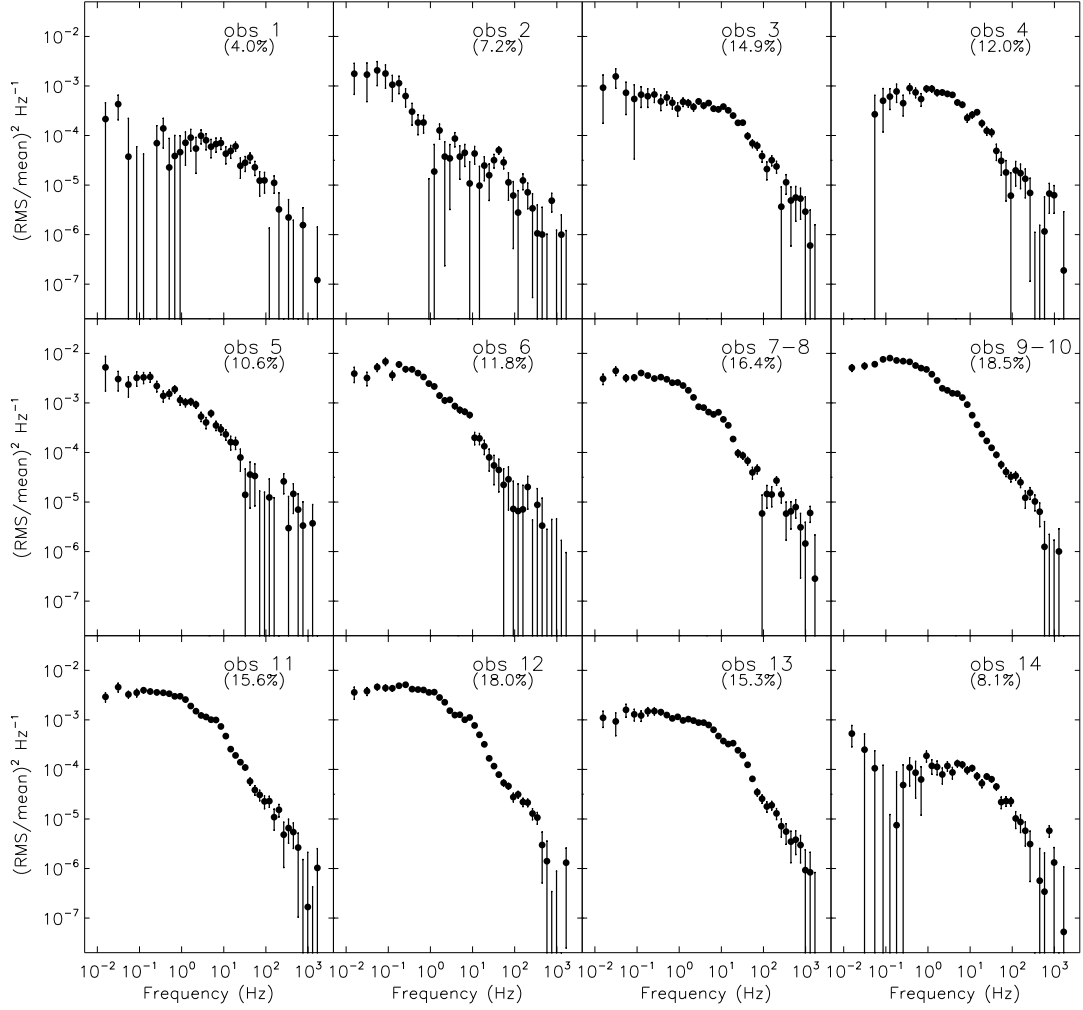}
\caption{Normalized Fourier power density spectra of \fouru. The
  integrated RMS value is given in each panel. Observations 7, 8 and
  9, 10 have very similar power density spectra and were combined in
  this figure. Note the presence around 10 Hz of a broad feature in observations 6 to 13 and the presence of noise extending above $\sim 500$
  Hz in observations 8 to 12. \label{fig10_bo}}
\end{center}
\end{figure*}

\begin{figure}[tb]
\begin{center}
\epsscale{1.0}
\plotone{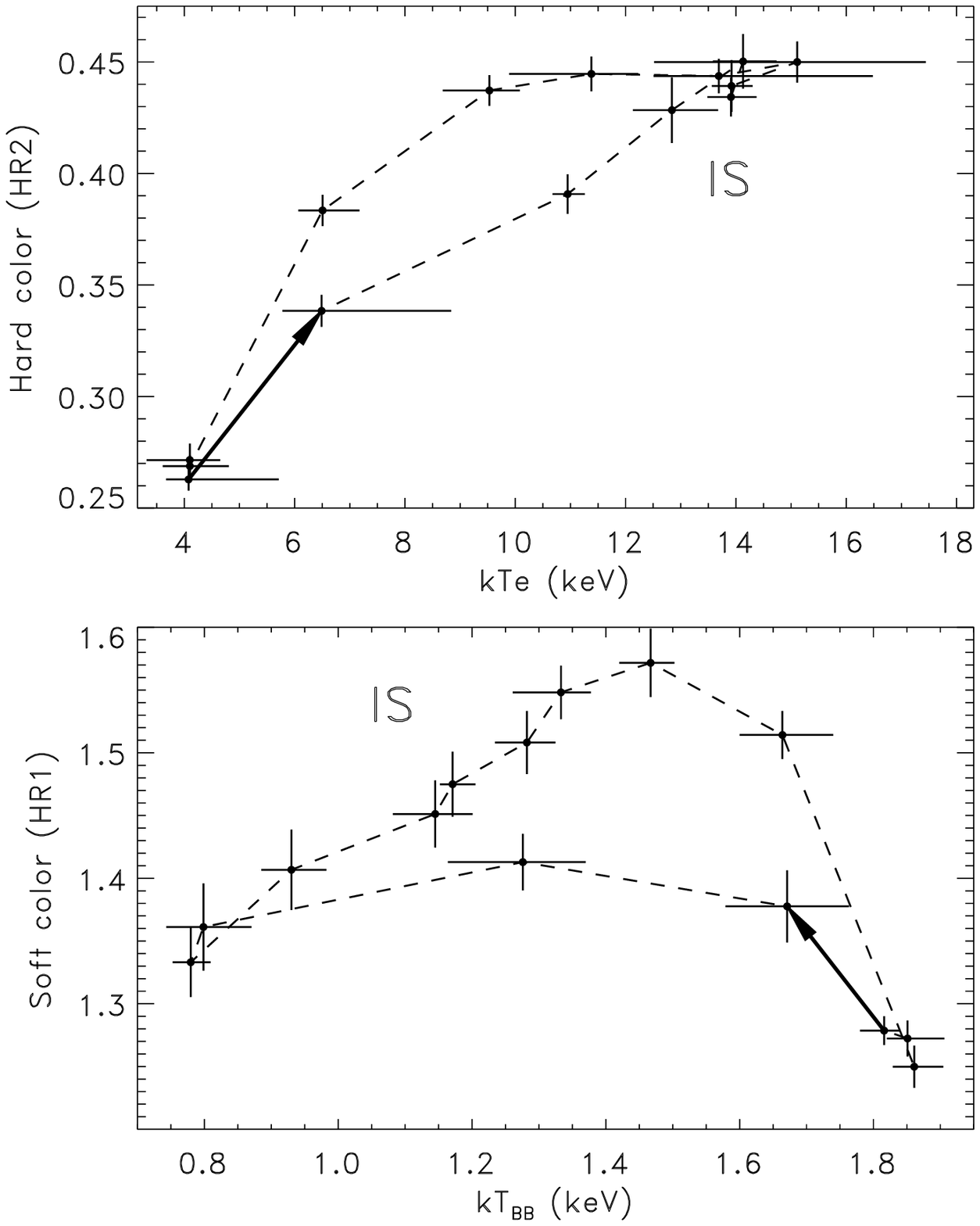}
\caption{The hard (top) and soft (bottom) colors as a
  function of the parameters to which they are the most sensitive to,
  namely the electron temperature (\kte)~and the blackbody temperature
  (\ktbb)~respectively. The arrows give the sense \fouru~moved on
  these plots. The soft color shows a very strong correlation with
  \ktbb~when the source is in the island state (IS).}
\label{fig11_bo}

\end{center}
\end{figure}

Comparing Figures \ref{fig1_bo} and \ref{fig10_bo}, the most significant difference is found between the diagonal branch power spectra of \fouru~and the normal branch power spectra of Z sources. On the normal branch, the power spectra of Z sources are characterized by power law noise at low frequencies below $\sim 1$ Hz, a normal branch oscillation ($\sim 5-7$ Hz) and above $\sim 10$ Hz a high frequency noise which cuts off between 50 and 100 Hz (Wijnands 2001,  see Fig \ref{fig1_bo}). The power spectra of \fouru~measured during the transition (e.g. observations 3, 4, 5 and 13) are different; they are characterized by band-limited noise with a strong high frequency component, and remain similar to those measured on the top branch of the Z (hard state). 

On the other hand, the power density spectra of \fouru~computed on the top branch of the Z (see Fig.  \ref{fig10_bo}) are at first order similar in shape (presence of band-limited noise, high frequency noise) with the horizontal branch power spectra of Z sources, as previously found (Wijnands \& van der Klis 1999; Wijnands 2001).  Note however that the feature seen around 10 Hz in observations 6 to 13 (see Fig. \ref{fig10_bo}), if similar to an horizontal branch oscillation, is less coherent in \fouru~than it is in Z sources (Wijnands 2001 and Fig \ref{fig1_bo}, see however Homan et al. 1998).  

Beside the possible difference in their power spectra on the diagonal branch, there are two other differences between Z sources and Atoll sources as illustrated in this paper; one lies in their energy spectra, the other one in the luminosities associated with the spectral transitions. The spectral changes of Z sources are much less spectacular than in Atolls (see e.g. Schultz \& Wijers 1993). In almost all states the energy spectra of Z sources remain very soft (see e.g. Di Salvo et al. 2000), resembling those of Atoll sources in their soft states (see Barret 2001 for a review of neutron star LMXB spectra). At the opposite, Atolls display large spectral variability with clear state transitions while moving on their CCD as
illustrated in this paper (see Figure \ref{fig7_bo}).  In Z sources,
the transitions are not related to large intensity changes (less than
a factor of 2) and take place at luminosities around
\ledd~(\ledd~is the Eddington limit assumed to be
  $2.5\times 10^{38}$\ergs, see e.g.  Schulz \& Wijers 1993; Di Salvo et al. 2000,
2001). In \fouru, the transitions occurred between 5 and 10\% \ledd,
and probably at lower luminosities in other sources (Gierli{\'n}ski \&
Done 2002a). 

Despite these differences, the observed similarities in the CCD of Atoll and Z sources suggest that the underlying geometries of these systems respond similarly to a varying mass accretion rate (Gierli{\'n}ski \& Done 2002a). Trying to accommodate the differences in the critical luminosity for state transitions and spectral shapes, Gierli{\'n}ski \& Done (2002a) proposed that they can be reconciled in a model where the only fundamental difference between Z and Atoll sources is the neutron star magnetic field. It is assumed that the magnetic field of Z sources is stronger than in Atolls (Lamb et al. 1985; Hasinger \& van der Klis 1989). The same spectral behavior can be observed in both Z and Atolls, but the higher accretion rate of Z implies that they are stable against the disk instability, and so vary only within a factor of $\sim 2$. In addition, the accretion rate is so high that evaporation is inefficient, and the truncation of the disk is caused by the neutron star magnetic field (Done 2002). Hence, the disk is truncated closer to the neutron star than in Atoll sources, the inner flow is thus cooler, giving rise to softer spectra (Gierli{\'n}ski \& Done 2002a). Because the accretion geometry remains basically the same, such a model could explain at the same time the similarities in the CCD shape of Atoll and Z sources, together with the differences in spectral shapes and in luminosities for state transitions. Now, it would be worth investigating whether it could also accommodate the difference in power spectra observed  between \fouru~and Z sources on the diagonal branch of the Z. This again will be discussed in a forthcoming paper.

\section{Conclusions}
We followed the spectral and timing evolution of \fouru~while it
moved on the upper branches of the Z reported recently in its
color-color diagram (Gierli{\'n}ski \& Done 2002a; Muno et al. 2002). The data set presented here is remarkable by the
fact that there are several observations with similar bolometric luminosities, but different properties of the energy and power density spectra. This nicely illustrates the difficulty in
interpreting snapshot observations of a variable source, for which the
long term history is unknown.  

We have related the changes in X-ray colors to changes in the energy spectrum modeled by the sum of a dominating comptonized component, a blackbody and a 6.4 keV line. The spectral transitions are primarily driven by changes of the electron temperature.  The spectral evolution of \fouru~can fit within a scenario in which the accretion geometry is made of a truncated accretion disk of varying radius and an inner flow merging smoothly with the neutron star boundary layer.  The soft component observed could be dominated by the accretion disk emission at the highest luminosities, and by the neutron star surface at the lowest luminosities. The comptonized component is always originating from the inner flow. The data presented here show that the truncation radius is not set by the instantaneous mass accretion rate, as observations with the same bolometric luminosity have very different spectral (and timing) properties.

Comparing the power density spectra of \fouru~and Z sources when they occupy similar branches of the Z shows that the most significant difference is on the diagonal branch on which the power density spectra of \fouru~remain similar to those measured on the top branch of the Z (the hard state). There are two other differences between Z sources and \fouru~both in the luminosity associated with the spectral transitions and on the spectral shape. 

\section{Acknowledgments}
This research has used the HEASARC database for retrieval of the
archival RXTE observations of \fouru. We are grateful  G. K.
Skinner, J. P. Lasota, J. M. Hameury for helpful comments. We thank M. Gierli{\'n}ski for a careful reading of the paper and for providing us with a paper on 4U1608-52 in advance of publication.

Finally, we wish to thank an anonymous referee for his/her excellent report which helped us to clarify some of the arguments presented in this paper.

\cleardoublepage

\begin{table}[t]
    \begin{center}
    \begin{tabular}{cccccc}

        \tableline
        Obs & Observation start & Observation end & T$_{\rm obs}$ (s) &
PCA (cts/s) & HEXTE-0 (cts/s) \\
        \tableline      
01 & 10/02/99-15:21:55 & 10/02/99-17:19:15 & 3648 & 481.7$\pm$ 7.2 &
6.5$\pm$ 2.5 \\
02 & 12/02/99-14:30:27 & 12/02/99-17:02:11 & 4368 & 400.2$\pm$ 7.3 &
5.5$\pm$ 2.2 \\
03 & 14/02/99-13:41:39 & 14/02/99-15:29:39 & 3552 & 258.6$\pm$ 5.5 &
7.6$\pm$ 3.0 \\
04 & 16/02/99-12:02:59 & 16/02/99-12:59:15 & 1584 & 175.8$\pm$ 4.1 &
8.1$\pm$ 2.5 \\
05 & 18/02/99-07:13:23 & 18/02/99-10:45:39 & 1776 & 109.3$\pm$ 3.3 &
6.4$\pm$ 2.9 \\
06 & 20/02/99-14:20:19 & 20/02/99-16:09:07 & 4288 &  94.5$\pm$ 3.2 &
5.8$\pm$ 2.4 \\
07 & 22/02/99-11:09:55 & 22/02/99-12:56:51 & 4288 & 131.9$\pm$ 3.4 &
6.8$\pm$ 2.2 \\
08 & 24/02/99-06:29:07 & 24/02/99-09:00:51 & 2720 & 174.8$\pm$ 5.3 &
10.2$\pm$ 2.9 \\
09 & 26/02/99-12:43:15 & 26/02/99-14:28:19 & 4448 & 195.4$\pm$ 4.8 &
11.9$\pm$ 2.7 \\
10 & 28/02/99-12:40:51 & 28/02/99-14:23:15 & 4112 & 241.9$\pm$ 6.4 &
14.8$\pm$ 2.4 \\
11 & 02/03/99-09:46:27 & 02/03/99-11:25:23 & 2496 & 320.5$\pm$ 7.8 &
19.8$\pm$ 3.3 \\
12 & 04/03/99-20:46:27 & 04/03/99-22:37:39 & 3568 & 382.8$\pm$ 7.0 &
23.1$\pm$ 3.2 \\
13 & 06/03/99-22:24:35 & 07/03/99-00:12:03 & 3808 & 450.5$\pm$12.1 &
18.0$\pm$ 3.0 \\
14 & 08/03/99-22:30:11 & 09/03/99-00:10:43 & 3472 & 571.6$\pm$ 9.2 &
8.9$\pm$ 2.3 \\
\tableline
\end{tabular}
     \caption{RXTE observation log of \fouru~in February-March 1999
(proposal P40051). The start and stop times of the observations are given
(from PCA files), together with the PCA exposure time, and the net source
count rate in the 3-16 keV range for PCA units 0, 1 and 2. The HEXTE-0 net
source count rate are also given in the 20-100 keV band (HEXTE-1 count
rates are not listed but they are about 25\% lower than the HEXTE-0 ones,
due to the loss of the pulse height analyzer of detector \#3).}
 \label{table1}
\end{center}
\end{table}

\small

\begin{table*}[t]

\begin{center}
    \begin{tabular}{ccccccccccccc}
        \tableline\tableline
        Obs & \ktbb & \rbb & \kte & $\tau$ & $\sigma_{\rm 6.4 keV}$ &
$\chi^2$/d.o.f & L$_{\rm Tot}$ & $f_{\rm BB}$ & EqW & $\eta$ & $f_{\rm
Hard}$   \\
\tableline

01 & $1.9^{+0.06}_{-0.03}$ & $ 1.8^{+ 0.1}_{- 0.3}$ & $ 4.1^{+ 0.6}_{-
0.8}$ & $7.1^{+ 1.6}_{- 0.7}$ & $0.63^{+ 0.13}_{- 0.14}$ & 50.6/68 & 25.6
& 19.3 & 126.0 &  0.0 &  1.2 \\
02 & $1.8^{+0.02}_{-0.04}$ & $ 1.8^{+ 0.1}_{- 0.1}$ & $ 4.1^{+ 1.6}_{-
0.4}$ & $7.0^{+ 0.8}_{- 1.4}$ & $0.63^{+ 0.13}_{- 0.12}$ & 54.5/68 & 21.1
& 20.6 & 134.0 & -0.8 &  1.1 \\
03 & $1.7^{+0.09}_{-0.09}$ & $ 1.0^{+ 0.1}_{- 0.1}$ & $ 6.5^{+ 2.3}_{-
0.7}$ & $6.7^{+ 0.6}_{- 1.3}$ & $0.81^{+ 0.10}_{- 0.11}$ & 77.6/69 & 13.9
&  7.7 & 273.0 &  1.2 &  7.7 \\
04 & $1.3^{+0.09}_{-0.11}$ & $ 1.3^{+ 0.3}_{- 0.2}$ & $10.9^{+ 0.3}_{-
0.3}$ & $5.5^{+ 0.0}_{- 0.0}$ & $0.79^{+ 0.17}_{- 0.18}$ & 131.5/117 &
10.7 &  5.4 & 236.0 & 12.6 & 19.2 \\
05 & $0.8^{+0.07}_{-0.06}$ & $ 3.0^{+ 0.7}_{- 0.8}$ & $12.8^{+ 0.8}_{-
0.7}$ & $5.5^{+ 0.0}_{- 0.0}$ & $0.93^{+ 0.23}_{- 0.25}$ & 82.5/117 &  7.5
&  6.6 & 279.0 & 22.0 & 26.6 \\
06 & $0.8^{+0.03}_{-0.03}$ & $ 3.5^{+ 0.4}_{- 0.4}$ & $14.1^{+ 0.6}_{-
0.5}$ & $5.5^{+ 0.0}_{- 0.0}$ & $0.70^{+ 0.17}_{- 0.13}$ & 116.2/117 &
6.9 &  8.7 & 217.0 & 27.0 & 31.0 \\
07 & $0.9^{+0.05}_{-0.04}$ & $ 2.5^{+ 0.4}_{- 0.3}$ & $13.9^{+ 0.5}_{-
0.4}$ & $5.5^{+ 0.0}_{- 0.0}$ & $0.86^{+ 0.11}_{- 0.15}$ & 97.5/116 &  9.3
&  6.6 & 258.0 & 24.5 & 30.8 \\
08 & $1.1^{+0.06}_{-0.06}$ & $ 1.7^{+ 0.2}_{- 0.2}$ & $13.9^{+ 0.4}_{-
0.4}$ & $5.5^{+ 0.0}_{- 0.0}$ & $0.63^{+ 0.16}_{- 0.16}$ & 98.5/111 & 12.3
&  5.4 & 156.0 & 24.2 & 31.3 \\
09 & $1.2^{+0.03}_{-0.02}$ & $ 1.8^{+ 0.1}_{- 0.1}$ & $15.1^{+ 2.3}_{-
2.6}$ & $5.5^{+ 0.7}_{- 0.5}$ & $0.51^{+ 0.11}_{- 0.11}$ & 90.4/110 & 14.6
&  5.5 & 131.0 & 28.6 & 35.7 \\
10 & $1.3^{+0.04}_{-0.05}$ & $ 1.7^{+ 0.1}_{- 0.1}$ & $13.7^{+ 2.8}_{-
1.2}$ & $5.8^{+ 0.4}_{- 0.6}$ & $0.70^{+ 0.15}_{- 0.16}$ & 130.9/110 &
17.5 &  5.5 & 155.0 & 26.4 & 34.2 \\
11 & $1.3^{+0.04}_{-0.07}$ & $ 1.7^{+ 0.2}_{- 0.1}$ & $11.4^{+ 0.8}_{-
1.5}$ & $6.7^{+ 0.6}_{- 0.3}$ & $0.89^{+ 0.13}_{- 0.12}$ & 110.2/119 &
22.3 &  5.6 & 225.0 & 23.5 & 32.0 \\
12 & $1.5^{+0.04}_{-0.05}$ & $ 1.8^{+ 0.2}_{- 0.2}$ & $ 9.5^{+ 0.6}_{-
0.8}$ & $7.4^{+ 0.5}_{- 0.3}$ & $0.75^{+ 0.16}_{- 0.16}$ & 119.8/110 &
24.7 &  7.7 & 177.0 & 17.4 & 27.7 \\
13 & $1.7^{+0.08}_{-0.06}$ & $ 1.6^{+ 0.1}_{- 0.1}$ & $ 6.5^{+ 0.7}_{-
0.4}$ & $7.9^{+ 0.4}_{- 0.5}$ & $0.75^{+ 0.10}_{- 0.10}$ & 93.1/72 & 24.1
& 10.0 & 218.0 &  0.4 & 12.4 \\
14 & $1.9^{+0.04}_{-0.03}$ & $ 1.8^{+ 0.1}_{- 0.2}$ & $ 4.1^{+ 0.7}_{-
0.5}$ & $7.0^{+ 1.0}_{- 0.9}$ & $0.67^{+ 0.11}_{- 0.10}$ & 70.6/68 & 31.1
& 16.6 & 174.0 &  2.0 &  1.2 \\
\tableline
\end{tabular}
    \caption{Best fit spectral parameters derived for \fouru. The
parameters are the blackbody temperature (\ktbb~in keV), the radius of the
blackbody computed at a distance of 7.4 kpc (\rbb~in km), the electron
temperature (\kte~in keV) and the optical depth of the spherical
scattering cloud as derived from the \comptt~model in XSPEC ($\tau$), the
width of the 6.4 keV line in keV ($\sigma_{\rm 6.4 keV}$), the $\chi^2$ of
the fit with the number of degree of freedom. All errors are given at the
90\% confidence level. L$_{\rm Tot}$ is the total 0.1-200 keV bolometric
luminosity is given in units of $10^{36}$ \ergs~at a distance of 7.4 kpc
(Haberl \& Titarchuk 1995). $f_{\rm BB}$ is the percentage of luminosity
carried out by the blackbody component. EqW is the equivalent width of the
6.4 keV line. $\eta$ is a bolometric correction factor. It gives in
percentage by which the luminosity would have been underestimated from a
direct scaling of the 3-16 keV count rates (normalized to observation 1).
$f_{\rm Hard}$ is the percentage of the source luminosity radiated above
20 keV.}
        \label{table2}
\end{center}

\end{table*}
\cleardoublepage

\begin{thebibliography}{}
\bibitem[Barret et al.(2000)]{ba00apj} Barret, D.,
Olive, J.~F., Boirin, L., Done, C., Skinner, G.~K., \& Grindlay, J.~E.\
2000, \apj, 533, 329
        
\bibitem[Barret(2001)]{ba01asr} Barret, D.\ 2001,
Advances in Space Research, 28, 307

\bibitem[]{}Belloni, T., Psaltis, D. \& van der Klis, M., 2002, ApJ,
  in press, astro-ph/0202213

\bibitem[]{}Bradt, H., Rothschild, R., E., Swank, J. H., 1993,
A\&AS, 97, 355
        
\bibitem[Di Salvo et al.(2001)]{2001ApJ...554...49D} Di Salvo, T.,
Robba, N.~R., Iaria, R., Stella, L., Burderi, L., \& Israel, G.~L.\ 2001,
\apj, 554, 49

\bibitem[Di Salvo et al.(2000)]{2000ApJ...544L.119D} Di Salvo,
T.~et al.\ 2000, \apjl, 544, L119

\bibitem[Done 2002]{do02}Done, C., 2002, submitted to Philosophical Transactions of the Royal Society (Series A: Mathematical, Physical, and Engineering Sciences), astro-ph/0203246 

\bibitem[Fender \& Hendry(2000)]{2000MNRAS.317....1F} Fender,
R.~P.~\& Hendry, M.~A.\ 2000, \mnras, 317, 1

\bibitem[Gierli{\'n}ski \& Done 2002a]{dg02mnras}Gierli{\'n}ski, M.
\& Done, C. 2002a, MNRAS, in press

\bibitem[Gierli{\'n}ski \& Done 2002b]{dg02mnras}Gierli{\'n}ski, M.
\& Done, C. 2002b, MNRAS, submitted
         
\bibitem[Hasinger \& van der Klis, 1989]{hv89aa}Hasinger \& van
der Klis, 1989, A\&A, 225, 79
        
\bibitem[Haberl \& Titarchuk(1995)]{1995A&A...299..414H} Haberl, F.~\&
Titarchuk, L.\ 1995, \aap, 299, 414

\bibitem[Homan et al.(1998)]{1998ApJ...499L..41H} Homan, J., van der Klis, 
M., Wijnands, R., Vaughan, B., \& Kuulkers, E.\ 1998, \apjl, 499, L41. 

\bibitem[Lamb, Shibazaki, Alpar, \& Shaham(1985)]{1985Natur.317..681L} 
Lamb, F.~K., Shibazaki, N., Alpar, M.~A., \& Shaham, J.\ 1985, \nat, 317, 
681. 
\bibitem[Langmeier et al.(1987)]{1987ApJ...323..288L} Langmeier, A.,
Sztajno, M., Hasinger, G., Truemper, J., \& Gottwald, M.\ 1987, \apj, 323,
288

\bibitem[Levine et al.(1996)]{1996ApJ...469L..33L} Levine, A.~M., Bradt,
H., Cui, W., Jernigan, J.~G., Morgan, E.~H., Remillard, R., Shirey, R.~E.,
\& Smith, D.~A.\ 1996, \apjl, 469, L33
        
\bibitem[Liu, van Paradijs, \& van den Heuvel(2001)]{2001A&A...368.1021L} Liu, Q.~Z., van Paradijs, J., \& van
den Heuvel, E.~P.~J.\ 2001, \aap, 368, 1021

\bibitem[M{\' e}ndez, 1999]{men99}M{\' e}ndez, M, 1989,
Proceedings of the Paris Texas Symposium, astro-ph/9903469

 \bibitem[]{mrc02apjl}Muno, M. P., Remillard, R., Chakrabarti, D.,
2002, \apjl, submitted
        
\bibitem[R{\' o}{\. z}a{\' n}ska \& Czerny(2000)]{2000MNRAS.316..473R}
R{\' o}{\. z}a{\' n}ska, A.~\& Czerny, B.\ 2000, \mnras, 316, 473

\bibitem[Schulz \& Wijers(1993)]{1993A&A...273..123S} Schulz,
N.~S.~\& Wijers, R.~A.~M.~J.\ 1993, \aap, 273, 123

\bibitem[Sunyaev \& Revnivtsev(2000)]{2000A&A...358..617S} Sunyaev, R.~\& 
Revnivtsev, M.\ 2000, \aap, 358, 617. 
\bibitem[Titarchuk(1994)]{1994ApJ...434..570T} Titarchuk, L.\ 1994, \apj,
434, 570

\bibitem[Tomsick, Corbel, \& Kaaret(2001)]{2001ApJ...563..229T} Tomsick,
J.~A., Corbel, S.~ \& Kaaret, P.\ 2001, \apj, 563, 229

\bibitem[van der Klis(2001)]{2001ApJ...561..943V} van der Klis, M.\ 2001, 
\apj, 561, 943. 
\bibitem[van Straaten et al.(2000)]{2000ApJ...540.1049V} van Straaten, S.,
Ford, E.~C., van der Klis, M., M{\' e}ndez, M., \& Kaaret, P.\ 2000, \apj,
540, 1049

\bibitem[Vrtilek et al.(1991)]{1991ApJS...76.1127V} Vrtilek, S.~D., 
McClintock, J.~E., Seward, F.~D., Kahn, S.~M., \& Wargelin, B.~J.\ 1991, 
\apjs, 76, 1127. 
\bibitem[Wijnands \& van der Klis 1999]{wv99apjl} Wijnands, R. \&
van der Klis, M., 1999, \apjl, 514, 939

\bibitem[Wijnands(2001)]{2001AdSpR..28..469W} Wijnands, R.\ 2001, Advances
in Space Research, 28, 469

\end{thebibliography}
\end{document}